\documentclass[fleqn,usenatbib]{mnras}

\usepackage{newtxtext,newtxmath}

\usepackage[english]{babel}
\usepackage[utf8]{inputenc}

\usepackage{ulem}  
\usepackage{color} 
\usepackage{amssymb}   
\usepackage{amsmath}   
\usepackage{bm}        
\usepackage[flushleft]{threeparttable}

\usepackage[pdftex]{graphicx}
\usepackage[outdir=./]{epstopdf}

\usepackage{rotating} 
\usepackage{caption}
\usepackage{booktabs}

\usepackage{hyperref}
\usepackage{cleveref}

\def\aap{A\& A}

\def\apj{ApJ}
\def\apjs{ApJS}
\def\apjl{ApJL}
\def\apss{Ap\& SS}

\def\araa{ARA\&A}

\def\mnras{{MNRAS}}

\def\pasj{{PASJ}}

\def\ssr{{SSRv}}

\newcommand{\beq}{\begin{equation}}
\newcommand{\eeq}{\end{equation}}

\newcommand{\Msun}{M_{\odot}}
\newcommand{\pc} {{\rm pc}}
\newcommand{\K}{{\rm K}}

\newcommand{\cmcub}{{\rm ~cm}^{3}}
\newcommand{\ergs}{{\rm erg} \, {\rm s}^{-1}}
\newcommand{\mG}{\mu {\rm G}}
\newcommand{\kms}{{\rm km~s}^{-1}}
\newcommand{\Linf}{L_{\rm inf}}
\newcommand{\LJ}{L_{\rm J}}
\newcommand{\Mach} {\mathcal{M}}
\newcommand{\Mc}{M_{\rm C}}
\newcommand{\cs}{c_{\rm s}}
\newcommand{\va}{v_{\rm A}}

\newcommand{\Minf} {\Mach_{\rm inf}}
\newcommand{\vinf} {v_{\rm inf}}
\newcommand{\Mrms} {\Mach_{\rm rms}}
\newcommand{\Ms}{M_{\rm S}}
\newcommand{\Msink}{M_{\rm Sink}}
\newcommand{\Msinks}{M_{\rm sinks}}
\newcommand{\mucrit}{\mu_{\rm crit}}
\newcommand{\Myr}{{\rm Myr}}
\newcommand{\ppcc}{{\rm cm}^{-3}}
\newcommand{\gpcc}{{\rm g ~cm}^{-3}}
\newcommand{\rhoth}{\rho_{\rm thr}}

\newcommand{\Rinf}{R_{\rm inf}}

\newcommand\redsout{\bgroup\markoverwith{\textcolor{red}{\rule[0.5ex]{2pt}{0.4pt}}}\ULon}

\title[Enhancement of the SFR by magnetic suppression of turbulence]{
	Magnetic Suppression of Turbulence and the Star Formation Activity of Molecular Clouds
}

\author[Zamora-Avil\'es et al. ]{Manuel~Zamora-Avil\'es,$^{1,2}$\thanks{E-mail: m.zamora@irya.unam.mx}  
	Enrique~V\'azquez-Semadeni,$^2$  Bastian~K{\"o}rtgen,$^3$ 
	\newauthor Robi~Banerjee,$^3$ and Lee~Hartmann$^1$ \\ 
	$^1$Department of Astronomy, University of Michigan, 311 West Hall, 1085 S. University Ann Arbor, MI  48109-1107 \\
	$^2$Instituto de Radioastronom\'ia y Astrof\'isica, UNAM. Apdo. Postal 72-3 (Xangari), Morelia, Michoc\'an 58089, M\'exico  \\
	$^3$Hamburger Sternwarte, Universit{\"a}t Hamburg, Gojenbergsweg 112, 21029 Hamburg, Germany}

\date{Accepted XXX. Received YYY; in original form ZZZ}

\pubyear{2017}

\begin{document}
\label{firstpage}
\pagerange{\pageref{firstpage}--\pageref{lastpage}}
\maketitle

\begin{abstract}
	
	We present magnetohydrodynamic simulations aimed at studying the effect of the magnetic suppression of turbulence (generated through various instabilities during the formation of molecular clouds by converging) on the subsequent star formation (SF) activity. We study four magnetically supercritical models with magnetic field strengths $B= 0$, 1, 2, and 3 $\mu$G (corresponding to mass--to--flux ratios of $\infty$, 4.76, 2.38, and 1.59 times the critical value), with the magnetic field, initially being aligned with the flows. We find that, for increasing magnetic field strength, the clouds formed tend to be more massive, denser, less turbulent, and with higher SF activity. This causes the onset of star formation activity in the non--magnetic or more weakly magnetized cases to be delayed by a few Myr in comparison to the more strongly magnetized cases. We attribute this behavior to the suppression of the nonlinear thin shell instability (NTSI) by the magnetic field, previously found by Heitsch and coworkers. This result is contrary to the standard notion that the magnetic field provides support to the clouds, thus reducing their star formation rate (SFR). However, our result is a completely nonlinear one, and could not be foreseen from simple linear considerations.
	
\end{abstract}


\section{Introduction}\label{sec:intro}

The effect of magnetic fields on the evolution of molecular clouds (MCs)
is not yet fully understood, in particular concerning its effect on the
formation of the clouds and their
subsequent star formation (SF) processes \citep[see e.g.,][]{Passot+95,HP00,HBB01,Ostriker+01,Kim+03,II08,II09,II12,Heitsch+09,Banerjee+09,VS+11,Inutsuka+15, Bastian-Robi15}.

One scenario that has been
recently proposed for the formation
of MCs (in solar neighborhood--like conditions) is that of
{\it colliding flows}, in which 
moderately supersonic streams of warm neutral
medium (WNM) collide to produce a cold neutral cloud through
nonlinear triggering of the thermal instability \citep[TI;][]{BP+99,HP99}. The cloud 
becomes turbulent through the combined action of TI and various 
other dynamical instabilities, such as the nonlinear thin shell (NTSI),
Kelvin-Helmholtz, and Rayleigh-Taylor instabilities
\citep[see e.g., ][]{Vishniac94, KI02, Heitsch+06, 
	VS+06}. Moreover, as a cloud forms and grows out of compression in the
WNM, it should become molecular, supercritical, and gravitationally
unstable at roughly the same time \citep[see e.g.,][]{HBB01}, transiting
to a regime of gravitational collapse. Because
this mechanism can coherently produce large amounts of cold, dense gas,
the newly formed clouds may contain numerous Jeans masses, and thus the
collapse can proceed in a hierarchical fashion, roughly as predicted by
\citet{Hoyle53}, but aided by the fact that the cloud's internal
turbulence produces nonlinear density fluctuations that have shorter
free-fall times than the cloud at large \citep{HH08, VS+09, GV14, VS15}.
This mechanism
provides a self-consistent way to generate the observed structure and
dynamics of MCs and to regulate the
evolution of the star formation rate (SFR) \citep[e.g.,][]{ZA+12, ZV14}.

However, as the magnetic field is ubiquitous in the Galaxy, it
is expected to play a central role in the formation of clouds. 
\cite{Hennebelle13} showed that regardless of the initial conditions 
(decaying turbulence or converging flows), the magnetic field increases the number 
of and builds more defined filamentary structures
due to the magnetic tension. These filaments also live longer. 

The ``magnetic criticality" is an important concept we will refer throughout the paper. To define it, we first define the mass--to--flux ratio (M2FR) as $\mu = \Sigma / \langle B_x \rangle$, where $\langle B_x \rangle$ is the mean magnetic field along the $x$--direction and $\Sigma$ is the surface density measured along the same direction and covering the entire length and cross section of the inflows \citep[see Sec. \ref{subsec:NS-initial-conditions}; see also][]{Banerjee+09, VS+11}. We then refer to magnetically supercritical/subcritical clouds if the M2FR is greater/lower than the critical value, $\mucrit = 0.16/\sqrt{G}$, which is an appropriate estimate for sheet-like clouds \citep{NN78} as those studied here.

In the case of converging flows, the strength and 
orientation of the mean magnetic field with respect to the flows play a
crucial role for the flow-driven formation of MCs, in particular for
whether the clouds can become dense, massive and supercritical
\citep[e.g.,] [] {Elm94, Passot+95, HP00, HBB01, II08, II09, II12, Heitsch+09,
	Banerjee+09, VS+11, HH14,Lazarian14, Bastian-Robi15}. This is simply because the gas
can flow freely along field lines, but in the perpendicular direction it
is opposed by the magnetic pressure, and because coherent compressions
are sometimes deemed difficult to maintain over the required scales of
hundreds of parsecs. Moreover, the magnetic tension 
tends to weaken or even suppress the dynamical instabilities (and
thus the generation of turbulence) in the clouds assembled by colliding
flows, particularly the NTSI \citep{Heitsch+07, Heitsch+09}
by preventing the transport of transverse momentum.

When the magnetic field is aligned with the WNM inflows, \cite{Heitsch+07} using pressure considerations and 2D simulations, found that the NTSI is suppressed if the Alfv\'en speed ($\va$) becomes comparable to the inflow velocity. Moreover, \cite{Heitsch+09}, using numerical simulations of colliding flows, found that both the strength and orientation of the magnetic field with respect to the inflows determine the morphology, thermal state, and the dynamics of MC precursors. Note that those authors did not include self-gravity in their analysis, and thus they did not study the resulting SF activity.

\cite{Banerjee+09} simulated the assembly of MCs by converging flows in a weakly magnetized medium including self--gravity. They studied the thermal and dynamical properties of clumps, finding that once the clumps are formed by the TI, they grow in mass and size by accretion from the WNM and eventually become Jeans unstable. They found that the  star formation efficiency (SFE) reaches $\sim50$\% at the end of the simulation.
\cite{VS+11} extended the latter work to explore higher magnetic field strengths (studying three models: one supercritical and two subcritical) and by taking into account diffusive processes, concluding that the SF activity is strongly attenuated in the subcritical cases by preventing the global collapse of the cloud. On the other hand, these authors reported an SFE of $\sim40$\% in the supercritical case.
Finally, \cite{Bastian-Robi15} presented a parameter study aimed at studying the transition from magnetically sub-- to super--critical states of cores immersed in subcritical clouds, finding that this is difficult to achieve even taking into account non--ideal MHD effects. Thus, the preferred mechanism for producing super-critical objects is by accretion from long distances, as described in \cite{HBB01} and \cite{VS+11}. 
However, note that, although the clouds studied by \citet{Bastian-Robi15} are globally supported by magnetic fields, they exhibit a 
strong turbulence suppression, particularly in models in which the inflows are aligned with the magnetic field lines (see their Fig. 12 and Table 1 below). This can be interpreted at the same time as an annihilation of the NTSI since the magnetic pressure overwhelmingly dominates the turbulent ram pressure \cite[as pointed out by ][]{Heitsch+07}.

In this contribution, we extend the previous works of \cite{Heitsch+07, Heitsch+09} by including self-gravity and modeling the SF process. We also extend the models by \cite{Banerjee+09,VS+11,Bastian-Robi15} by studying four magnetically supercritical clouds, focusing in the effect of magnetic suppression of turbulence on the SF activity (see Table \ref{tab:pwork} for a comparison of the initial conditions used in each work).

Finally, the plan of the paper is as follows: In Section \ref{sec:inestabilities} we briefly describe the main instabilities occurring during the formation of MCs by converging flows. In Section \ref{sec:simulations} we
describe the numerical model. The results are presented in Sec.\
\ref{sec:results} and discussed in Sec.\ \ref{sec:discussion}. Finally, 
a summary and some conclusions are presented in Section
\ref{sec:conclusions}.

\begin{table*}
	\centering
	\caption{Overview of the initial conditions used in works with similar setup, i.e., two WNM colliding flows aligned with the magnetic field (see Sec. \ref{subsec:NS-initial-conditions} for a detailed description). Here, $n_0$ is the initial number density, $\cs$ the isothermal sound speed of the WNM, $\vinf$ the inflow velocity, $\Rinf$ is half of the inflow diameter (the inflow length, $\Linf$, is 112 pc in all works, except in Heitsch et al. (2009) since they consider inflow boundary conditions), $\Mrms$ is the background turbulence, $|B_0|$ is the initial magnetic field strength, $\mu$ the mass--to--flux ratio (M2FR), and $\va$ the alfv\'en speed ($= B_0/ \sqrt{4 \pi \rho}$).}
	\label{tab:pwork}
	\begin{threeparttable}
		\resizebox{\textwidth}{!}{
			\begin{tabular}{cccccccccc}
				\hline
				& $n_0$   & $\cs$  & $\vinf$  & $\Rinf$ &       &$|B_0|$    &                             &     \\ 
				&($\ppcc$)&($\kms$)& ($\kms$) & (pc)    &$\Mrms$&($\mG$)    &$\mu/\mucrit$\tnote{a}       & $\va / \vinf$\tnote{b} \\
				\hline
				\cite{Heitsch+09}            &  1      & 7.2    & 16.0     & 22      & -     &0,2.5,5.0  &-                            &0,0.3,0.6  \\[2pt]
				\cite{Banerjee+09}           &  1      & 5.7    & 7.1      & 32      & 0.12  &1          &\boldsymbol{$2.37$}          &0.27   \\[2pt]
				\cite{VS+11}                 &  1      & 5.7    &  13.9    & 32      & 0.12  &2,3,4      &\boldsymbol{$1.36$},0.91,0.68         &0.27,0.42,0.56   \\[2pt]
				\cite{Bastian-Robi15}        &  1     & 5.7    &  11.4    & 64      &0.4,0.4,0.5\tnote{c}&3,4,5\tnote{c}               &0.79,0.59,0.47\tnote{c}       &0.5,0.67,0.84\tnote{c}   \\[2pt]
				This work                    &  2     &  3.1   &  7.5     & 32      & 0.7   &0,1,2,3     &\boldsymbol{$\infty, 4.76,2.38,1.59$}\tnote{d}&0,0.18,0.36,0.54\tnote{b}   \\
				\hline
			\end{tabular} }
			\begin{tablenotes}\footnotesize
				\small
				\item[a] The bold numbers correspond to magnetically supercritical models.
				\item[b] \cite{Heitsch+07} found that the NTSI is supressed when $\va / \vinf \gtrsim 1$ (i.e., when the magnetic pressure overcomes the tangential component of the ram pressure). According to this criterion we expect the development of the NTSI in all the models considered here (see Table \ref{table:sims}).
				\item[c] These models are labeled as B3M0.4I0, B4M0.4I0, and B5M0.5I0, respectively, in \cite{Bastian-Robi15}.
				\item[d] Note that in this work, the initial density is $n_0=2$ cm$^{-3}$, which increases the magnetic criticality
				($\mu/\mucrit$) by a factor of 2 with respect to previous works using instead $n_0=1$ cm$^{-3}$ (but the same magnetic field strength and inflow dimensions).
			\end{tablenotes}
		\end{threeparttable}
	\end{table*}

	\section{Overview of instabilities} \label{sec:inestabilities}
	
	Several instabilities enable the condensation and fragmentation of clouds formed by colliding flows, driving turbulent motions in the compressed, cooled layer. Here we briefly review the basics of the thermal, nonlinear thin-shell, Kelvin-Helmholtz and Jeans instabilities. For a more detailed discussion see e.g., \cite{Heitsch+06,Heitsch+08,VS15} and references therein.
	
	\subsection{Nonlinear Thin-Shell Instability (NTSI)} \label{subsec:NTSI}
	
	In the hydrodynamic (HD) limit, the NTSI \citep{Vishniac94} is triggered in a bent cold slab formed by two accretion shocks when the slab displacement, $\eta$, is comparable to its thickness. The growth rate is $c_{\rm s}k(k \eta)^{1/2}$, where $k$ is the wave number of the slab perturbation and $c_{\rm s}$ the sound speed.
	
	The NTSI acts by increasing the curvature of bending perturbations on the boundary of the dense compressed layer and pushing the gas towards the nodes of these perturbations. The gas therefore moves in a direction oblique to the inflows, effectively transferring part of the momentum from the inflows to the direction perpendicular to them \citep{Vishniac94}. The oblique convergence of gas at opposite sides of the nodes finally results in a compression and accele\-ra\-tion of the gas in the nodes parallel to the inflows, but moving in the opposite direction, which ultimately produces an expansion of the layer \citep[e.g., ][]{FW06}.
	
	On the other hand, in the MHD regime it is expected that magnetic fields soften the NTSI via the magnetic pressure term of the Lorentz force. Thus, in the configuration studied here, when the magnetic field is aligned with the inflows, the magnetic tension force prevents motions perpendicular to the magnetic field and it can even suppress the NTSI when the Alfv\'en speed becomes comparable to the inflow velocity \citep{Heitsch+08}. 
	
	\subsection{Kelvin-Helmholtz Instability (KHI)} \label{subsec:KHI}
	
	The KHI \citep[see e.g.,][]{Chandrasekhar61} is a velocity-shear insta\-bi\-li\-ty, which is also expected to appear in WNM converging flows as an important turbulence driver in the MC precursors.
	
	In the compressible HD regime, when the velocity shear is in a single continuous fluid, the instability growth rate is given by $k \Delta U$, where $k$ is the wave number of the perturbation and $\Delta U$ is the velocity jump across the shear layer. In the MHD limit, when the magnetic field is uniform and parallel to the shear flow, the magnetic tension force can reduce or even stabilize the growth rate when the velocity jump is less than the Alfv\'en speed \cite[][and references therein]{Heitsch+06}.

	\subsection{Thermal Instability (TI)} \label{subsec:TI}
	
	This instability is closely related to the cooling and heating processes. The functional form of the cooling function (which we discuss in Sec. \ref{sec:heating-cooling}) causes an atypical thermodinamical behavior in the ISM known as TI \citep{Field65}. This instability can be understood as the tendency of the ISM to depart from the thermal equilibrium condition upon a perturbation, thus producing condensations of cold and dense regions immersed in a WNM.\footnote{This instability can proceed in two regimes, the isochoric and isobaric modes. Here we focus on the isobaric mode, which is the relevant one in the pc-scales we are interested in (see, e.g., the review by \cite{VS+03}).} 
	In the linear regime, the condensations can survive at scales ($\lambda$) smaller than the sound crossing length $\lambda_{\rm s}=\tau_{\rm c} c_{\rm s}$, $\tau_c$ being the cooling timescale and $c_{\rm s}$ the adiabatic sound speed.
	The cooling timescale is:

	\beq 
	\tau_c = \frac{3}{2} \frac{k_{\rm B} T}{|\Gamma - n \Lambda|}, 
	\eeq
	where $k_{\rm B}$ is the Boltzmann constant, $T$ the temperature, and $\Gamma$ and $\Lambda$ the heating and cooling functions, respectively.
	This instability is important in early stages of cloud formation since it is an efficient mechanism to generate non linear density perturbations and turbulence on small scales. 
	Note, however, that turbulence can affect the behavior of the TI at pc-scales. In this regard, \cite{SS+02} have shown that if the velocity fluctuations are dominated by the turbulence rather than by the adiabatic sound speed the growth of the perturbations can be suppressed even though $\lambda < \lambda_{\rm s}$.
	
	\subsection{Jeans instability} \label{subsec:JI}
	
	This well known instability occurs in a self-gravitating uniform medium (with density $\rho_0$), in the presence of thermal pressure, to which linear sinusoidal perturbations are added. In this case, pertubations of wavelength larger than the so-called Jeans length \citep{Jeans1902}
	\beq \label{eq:lJ}
	\lambda_{\rm J} = \Big( \frac{\pi c^2_{\rm s}}{G \rho_0} \Big),
	\eeq
	cannot be supported by the pressure gradient and proceed to collapse.
	
	\section{Numerical model} \label{sec:simulations}
	
	\subsection{Numerical scheme} \label{sec:scheme}
	
	We use the Eulerian adaptive mesh refinement {\tt FLASH}2.5 code
	\citep{Fryxell+00} to perform three-dimensional, self gravitating, \textrm{idela} MHD
	simulations, including heating and cooling processes. 
	The ideal MHD
	equations are solved using the MHD HLL3R solver, which
	preserves positive states for density and internal energy
	\citep{Bouchut+07,Bouchut+10,Waagan09, Waagan+11}. This solver is
	suitable for highly supersonic astrophysical problems, such as those
	studied here.

	\subsection{Heating and Cooling} \label{sec:heating-cooling}
	
	We calculate the heating and cooling rates using the analytic fits by
	\citet{KI02} for the heating ($\Gamma_{\rm KI}$) and cooling
	($\Lambda_{\rm KI}$) functions \citep[see also][for corrections to
	typographical errors]{VS+07},
	\beq
	\Gamma_{\rm KI} = 2.0 \times 10^{-26}  \, \ergs,
	\eeq
	\begin{multline}
	\frac{\Lambda_{\rm KI}(T)}{\Gamma_{\rm KI}} = 10^7 \exp \Big( {\frac{-1.184
			\times 10^5}{T+1000}} \Big) + \\ + 1.4 \times 10^{-2} \sqrt{T} \exp \Big( {\frac{-92}{T}} \Big) \cmcub,
	\end{multline}
	which are based on  the thermal and chemical calculations considered by
	\citet{Wolfire+95} and \citet{KI00}. 
	
	In the present study, we have chosen {\it not} to include any form
	of stellar feedback, in order to isolate the effects of varying the
	magnetic field. A study including stellar feedback will be
	presented elsewhere.

	\subsection{Refinement criterion and sink particles} \label{subsec:sinks}
	
	In order to follow the
	development of high density regions, we employ a refinement
	criterion which, however, is not standard. The most frequently used
	criterion for refinement is the so-called Jeans criterion
	\citep{Truelove+97}, which aims to prevent artificial fragmentation,
	requiring that the local Jeans length be
	resolved by at least a certain minimum of grid cells. Because, at
	constant temperature, the Jeans length scales as $\rho^{-1/2}$, the
	condition of a fixed number of cells per Jeans length implies that the
	grid size also scales with density as $\Delta x \propto \rho^{-1/2}$. Instead, we have used a {\it constant mass} criterion,
	in which the grid size scales with density as $\Delta x \propto
	\rho^{-1/3}$, and so the mass of the new level of refined cells is the
	same as that of the previous level when it was created.\footnote{The reason
		for this is that the present simulations are part of a larger set (to be
		presented elsewhere) that contains also simulations with stellar
		feedback, where the feedback is a function of the sink particle mass. In
		those simulations, the priority is for the
		feedback to be independent of
		resolution, and such a refinement criterion achieves this goal.}
	In section \ref{subsec:ref} we show that the effect in 
	using the {\it constant mass} (in run B3)
	instead of the Jeans criterion (run B3J) is just a delay in the onset of star 
	formation by $\sim 10$\%. 
	Therefore, we do not consider that the usage of this 
	criterion has any important consequences, since
	we are not concerned with the mass distribution of the sinks (related to
	the fragmentation), but only with the net SFR.
	
	Once the maximum refinement level is reached in a given cell, a sink particle can be formed when the
	density in this cell exceeds a threshold number density, 
	$\rhoth = 8.9 \times 10^{-18} \, \gpcc$ ($=4.2 \times 10^{6} \,
	\ppcc$), among other sink-formation
	tests \citep[for a detailed description and the implementation in the FLASH code, see][]{Federrath+10}. The sink is
	formed with the excess mass within the cell, that is,
	$\Msink=(\rho-\rhoth) \Delta x^3$. The sink particles can then accrete
	mass (with $\rho > \rhoth$) from their surroundings \citep[within an accretion
	radius of $\sim 2.5 \Delta x$;][]{Federrath+10}.

	\subsection{Initial conditions} \label{subsec:NS-initial-conditions}
	
	We use a setup similar to that of \citet{VS+07}. The numerical periodic
	box, of dimensions $L_x=256 \, \pc$ and $L_y = L_z
	=128 \, \pc$, is initially filled with warm neutral gas at a uniform
	density $n = 2 \, \ppcc$ and constant temperature of $1450 \,
	\K$, 
	which corresponds to
	thermal equilibrium and implies an isothermal sound speed
	of 3.1 $\kms$.  Assuming a composition of atomic
	hydrogen only (with a mean molecular weight $\mu = 1.27$), the gas mass
	in the whole box is $\sim 2.6 \times 10^5 \, \Msun$, whereas the mass
	contained in the cylinders is $\sim 4.5 \times 10^4 \, \Msun$.
	
	In order to trigger the NTSI, we impose an initial turbulent velocity field, which corresponds to a Burgers turbulence, with a power spectrum of  $k^{-2}$ and 
	Mach number of $\Mrms \simeq$0.7. No forcing is applied at later times and therefore the turbulence decays.
	On top of this
	weak background turbulent field, we add two cylindrical streams, each of radius $\Rinf=32
	\, \pc$ and length $\Linf = 112 \, \pc$, moving in opposite directions
	at a moderately supersonic velocity of $7.5 \, \kms$ in the
	$x$-direction (implying a Mach number $\Minf \simeq$ 2.4 with
	respect to the isothermal sound speed in the WNM).
	
	The numerical box is permeated with a uniform magnetic field along the
	$x$-direction, for which we consider three different
	strengths ($1$, $2$, and $3 \, \mG$), along with a nonmagnetic case.
	Note that the corresponding 
	mass-\-to-\-flux ratios are greater than the critical value, so that our
	clouds are mag\-ne\-ti\-cally supercritical in all cases, in
	agreement with observations that suggest that MCs are in general
	magnetically supercritical \citep{Crutcher+10}. Table
	\ref{table:sims} gives a summary of the magnetic parameters of the
	simulations. Finally, the maximum resolution reached is $\Delta x
	= 0.03 \, \pc$.
	
	With this setup and initial conditions we perform four (otherwise
	identical) simulations, varying only the magnetic field strength. We
	only consider the case where the magnetic field is
	initially aligned with the inflows (along the $x$ direction), but
	note that non-aligned cases can lead to motions along the magnetic field
	for sufficiently high inflow speeds at a given field strength
	\citep[see e.g.,][]{HP00,Bastian-Robi15}. 
	
	\begin{table}
		\caption{List of simulations}
		\label{table:sims}
		\begin{threeparttable}
			\begin{tabular}{c c c c}
				\hline\hline
				Run name & $|{\rm B}_{x,0}| \, (\mG)$ & $\mu / \mucrit$\tnote{a} & Refinement criterion\tnote{b}  \\
				\hline
				B0 & $0$ & $\infty$ & constant mass   \\
				B1 & $1$ & $4.76$   & constant mass \\
				B2 & $2$ & $2.38$   & constant mass \\
				B3 & $3$ & $1.59$   & constant mass \\
				B3J & $3$ & $1.59$  & Jeans\\ 
				\hline 
			\end{tabular}
			\begin{tablenotes}\footnotesize
				\small
				\item[a] Note that all the models are magnetically supercritical.
				\item[b] See Section \ref{subsec:ref} for a discussion about the refinement criteria.
			\end{tablenotes}
		\end{threeparttable}
	\end{table}
	
	\section{Results}\label{sec:results}
	
	Henceforth, the analysis focuses on the ``central box'' of the
	simulation, a cylindrical region of radius 40 pc and length of
	$80 \, \pc$, centered at the plane where the
	flows collide. We will refer to the cold cloud formed in this box as the
	``central cloud'', or simply, ``the cloud'', and here we will
	consider its global physical properties. 
	
	
	\begin{figure*}
		\hfill
		\includegraphics[height=23.3cm,keepaspectratio]{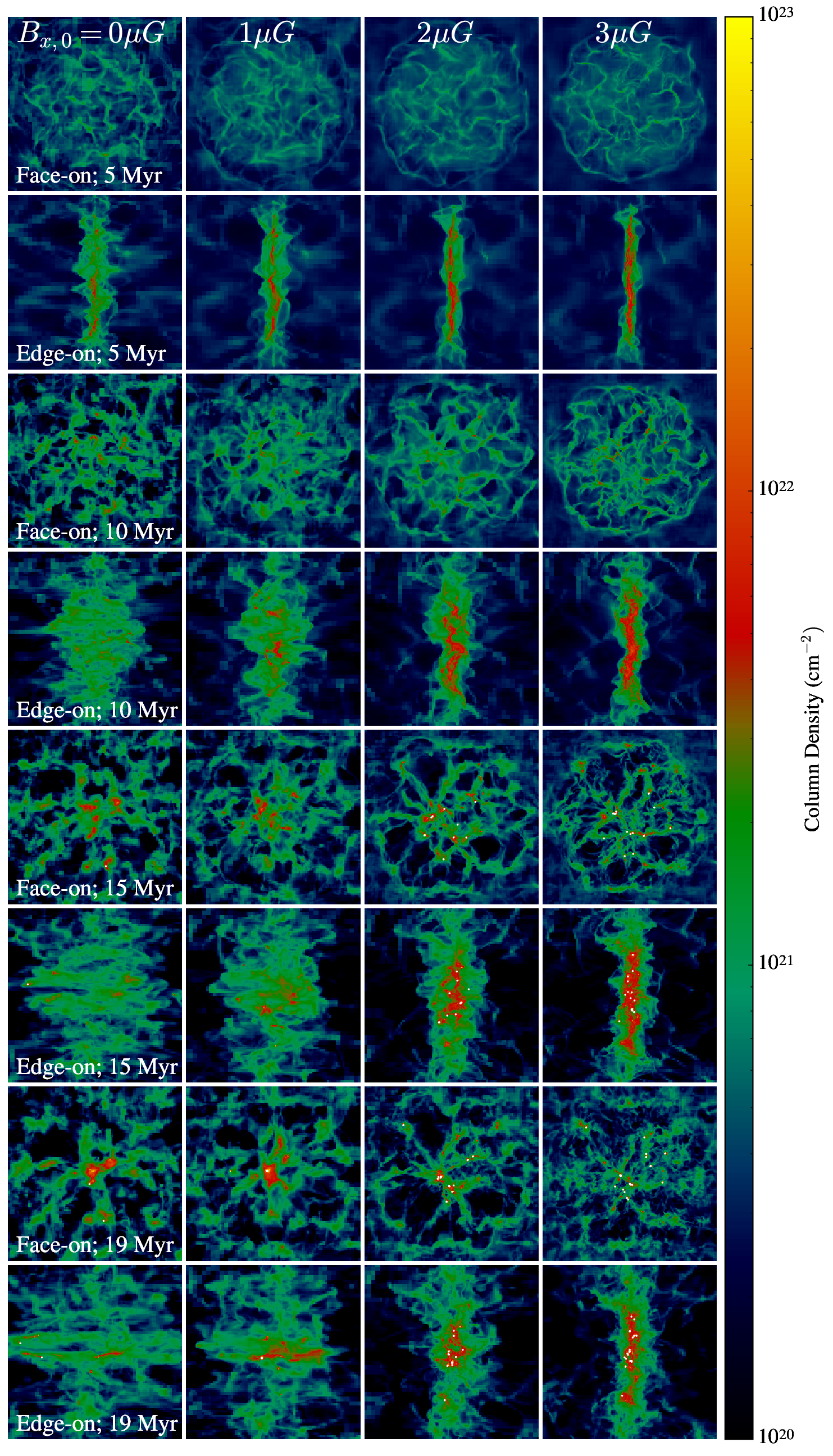}
		\hspace{\abovecaptionskip}
		\begin{minipage}[b][23cm][c]{2\baselineskip}
			\rotcaption{\label{fig:cdens} Face-on and edge-on column density views of the ``central clouds'' at $t=5$, $10$, $15$, and
				$19 \, \Myr$. The different columns represent the models B0 ($B_{x,0}=0 \,
				\mG$), B1 ($1 \, \mG$), B2 ($2 \, \mG$), and B3 ($3 \, \mG$). The dots (in panels at $t=15$ and $19 \, \Myr$) represent the 
				projected position of the sink particles, i.e., collapsed objects. The box
				is $80 \, \pc$ per side. Note that in the simulations with
				larger magnetic field strengths, the clouds are more coherent and
				compact.}
		\end{minipage}
		\hfill\mbox{}
	\end{figure*}
	%
	
	\subsection{Global evolution} \label{subsec:glob_evol}
	
	The collision of WNM streams (or {\it inflows}) in the center of the
	numerical box nonlinearly triggers thermal instability (TI),
	forming a thin cloud of cold atomic gas \citep[see
	e.g.,][]{HP99,KI00,KI02,WF00}, which at the same time becomes turbulent
	by the combined action of TI with various other dynamical
	instabilities \citep[see e.g.,][]{Hunter+86,Vishniac94, KI02,
		Heitsch+05, VS+06}. Also, the total (thermal plus
	ram) pressure in the dense gas is in
	balance with the total pressure of
	the surrounding WNM, while the thermal pressures are also equal because
	of the rapid condensation caused by TI. It was shown by
	\citet{Banerjee+09} that the two conditions combined imply similar
	turbulent Mach numbers in both the diffuse and the dense gas, and so the
	cold, dense gas exhibits a moderately supersonic velocity dispersion.
	Since we do not follow the chemistry we
	indistinctly will refer to the dense gas ($n>100 \, \ppcc$) as 
	``molecular''. 
	
	
	\begin{figure}
		\includegraphics[width=1\hsize]{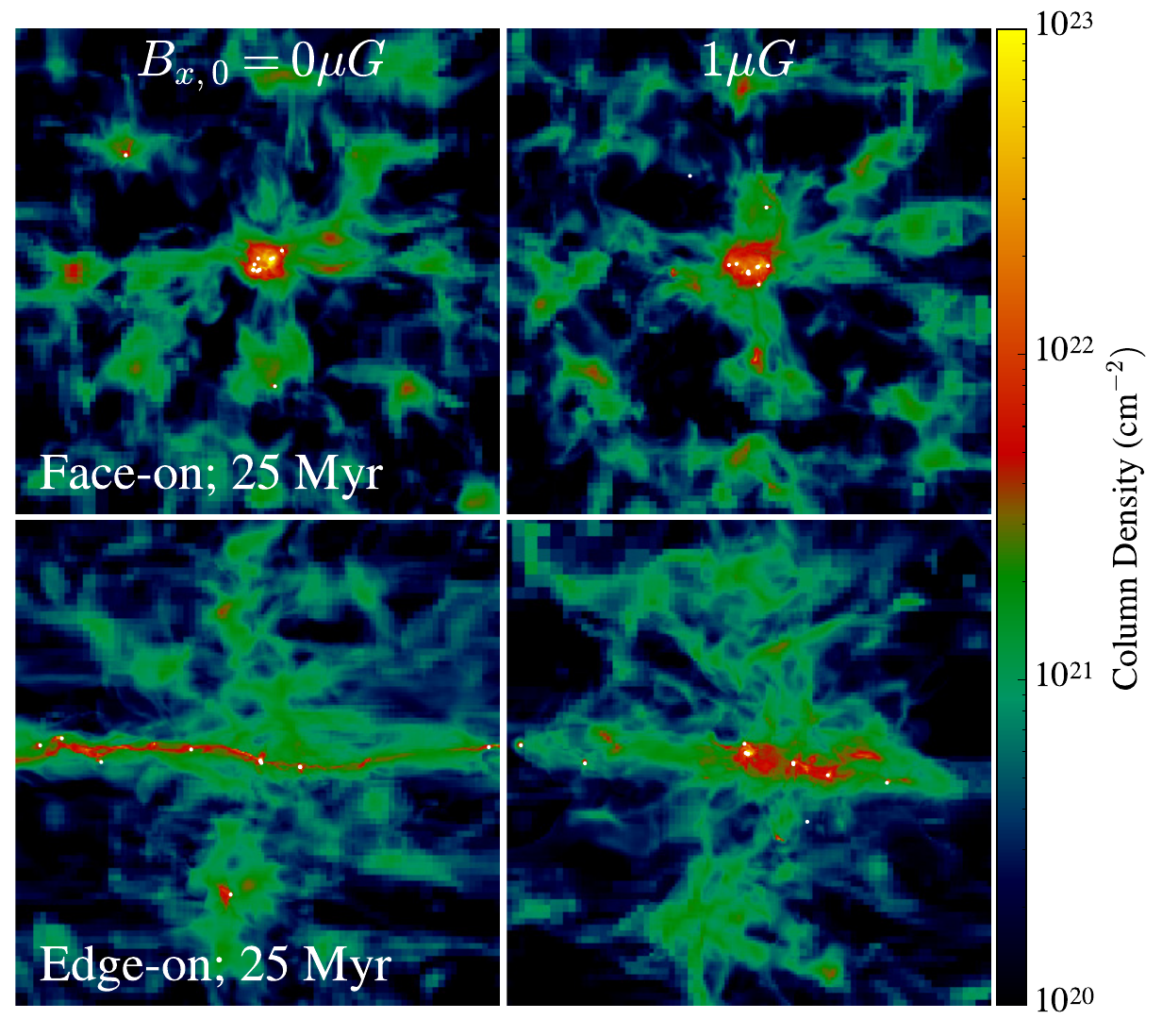}
		\caption{\label{fig:cdens25} Same as Fig.\ \ref{fig:cdens}, but at
			$t=25 \, \Myr$. By this time, runs B2 and B3 have already been terminated.}
	\end{figure}
	
	Since our numerical
	boxes are magnetically supercritical (see Table \ref{table:sims}),
	the clouds soon become 
	dominated by self-gravity\footnote{As \citet{HBB01, VS+11,HH14} have pointed out, MCs are expected to start their lives as magnetically subcritical, cold atomic clouds, and evolve towards being mostly molecular and magnetically supercritical as they accrete material along the magnetic field lines.}
	and begin to contract gravitationally as a whole. During the
	large-scale contraction, the local, nonlinear (i.e., large-amplitude)
	density enhancements produced by the initial turbulence manage to
	complete a collapse of their own, since their local free-fall time is
	shorter than the average one for the entire cloud \citep{VS+07,VS+09, HH08,
		Pon+11}. These local collapses involve only a small fraction of the
	cloud's total mass (see Sec.\ \ref{sec:sfe-sfr}).
	
	
	\subsection{Cloud structure}\label{subsec:structure}
	
	Figures \ref{fig:cdens} and \ref{fig:cdens25} illustrate 
	the morphological effects of magnetic fields at $t=5$, 10, 15, 19, and
	25 Myr. In Fig. \ref{fig:cdens} we show $t=19$ Myr
	rather than 20 Myr in order to show all four simulations, since run B2
	terminates before this time.\footnote{We evolve the
		simulations for $\sim$8-10 Myr after the onset of SF since this period is large enough to
		observe a trend in the SF activity in each simulation.}
	
	In these figures, the various columns from left to
	right represent column density maps of the ``central clouds'' for models
	B0, B1, B2, and B3 ($B_{x,0}=0$, $1$, $2$, and
	$3 \, \mG$, respectively) in face-on and edge-on views. The morphologies (in column density maps) are consistent
	with those reported by \citet[][see their models H, X25, and X50 in
	their Figure 1]{Heitsch+09}.
	
	From these figures, it is seen that {\it increasing the magnetic
		field strength prevents the cloud from reexpanding in the direction of
		the inflows, as well as from growing in the direction
		perpendicular to them} after they collide. 
	This can be interpreted as an
	inhibition of the NTSI by the magnetic field \citep[see also,][]{Heitsch+07,Heitsch+09}.
	We speculate that this occurs because the magnetic field, which is
	parallel to the inflows, inhibits the transverse motions that would
	develop as a consequence of the NTSI,
	and therefore allows
	for more efficient dissipation of the inflow kinetic energy by forcing a
	more focused collision of the two streams.
	
	\begin{figure}
		\includegraphics[width=1\hsize]{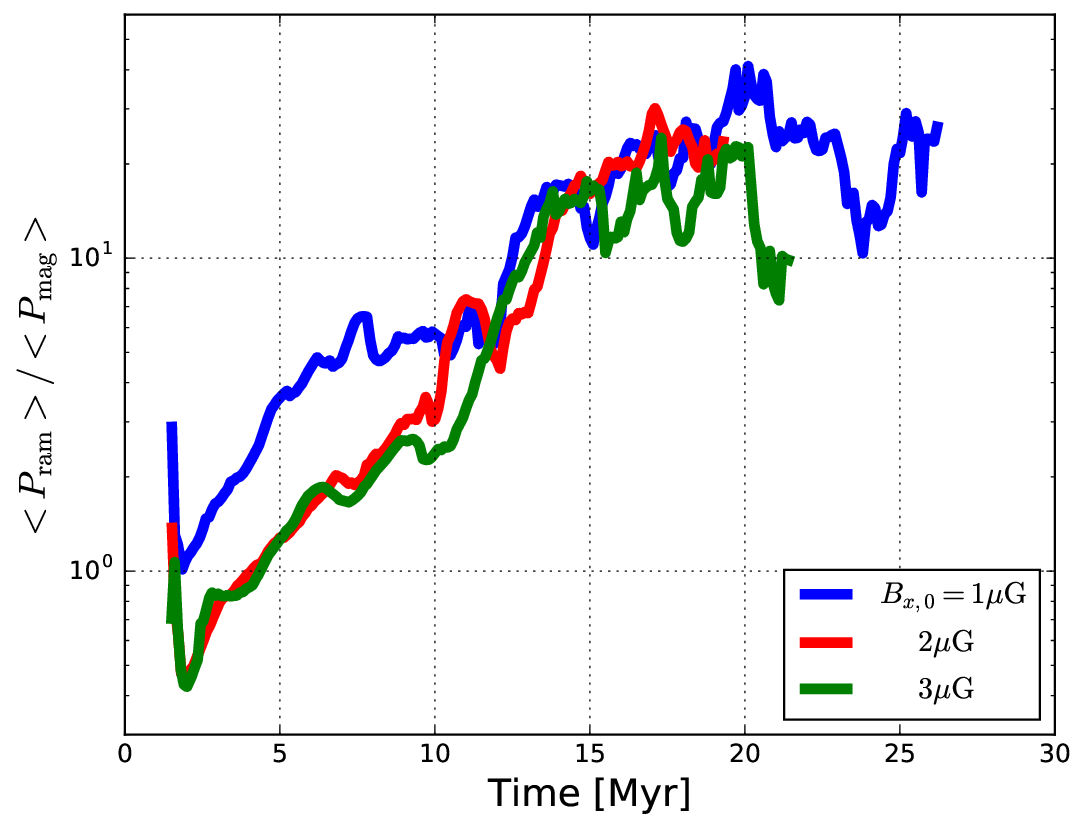}
		\caption{Time evolution of the ratio of ram to magnetic pressure (arithmetically averaged).}
		\label{fig:pram_pmag}
	\end{figure}
	
	To test for this speculation, in Fig. \ref{fig:pram_pmag} we plot the arit\-me\-tic average of the ram to magnetic pressure ratio ($\beta_{\rm ram} \equiv P_{\rm ram} /  P_{\rm mag}$) in the dense slab ($n \gtrsim 100 \, \ppcc$) for models B1, B2, and B3. 
	Based on simple pressure arguments, we expect the development of the NTSI roughly when $\beta_{\rm ram} > 1$ \citep{Heitsch+07}. Thus, for model B1 the NTSI operates readily once the slab forms, whereas for models B2 and B3 the instability starts to act after $t \sim 5 \, \Myr$. Interestingly, and regardless of the model, the ram pressure increases by roughly three orders of magnitude throughout the cloud evolution, whereas the magnetic pressure increases smoothly by only one order of magnitude, resulting in a net growth of the $\beta_{\rm ram}$ parameter, as shown in Fig. \ref{fig:pram_pmag}.
	
	%
	
\begin{figure*}
		\includegraphics[width=0.49\hsize]{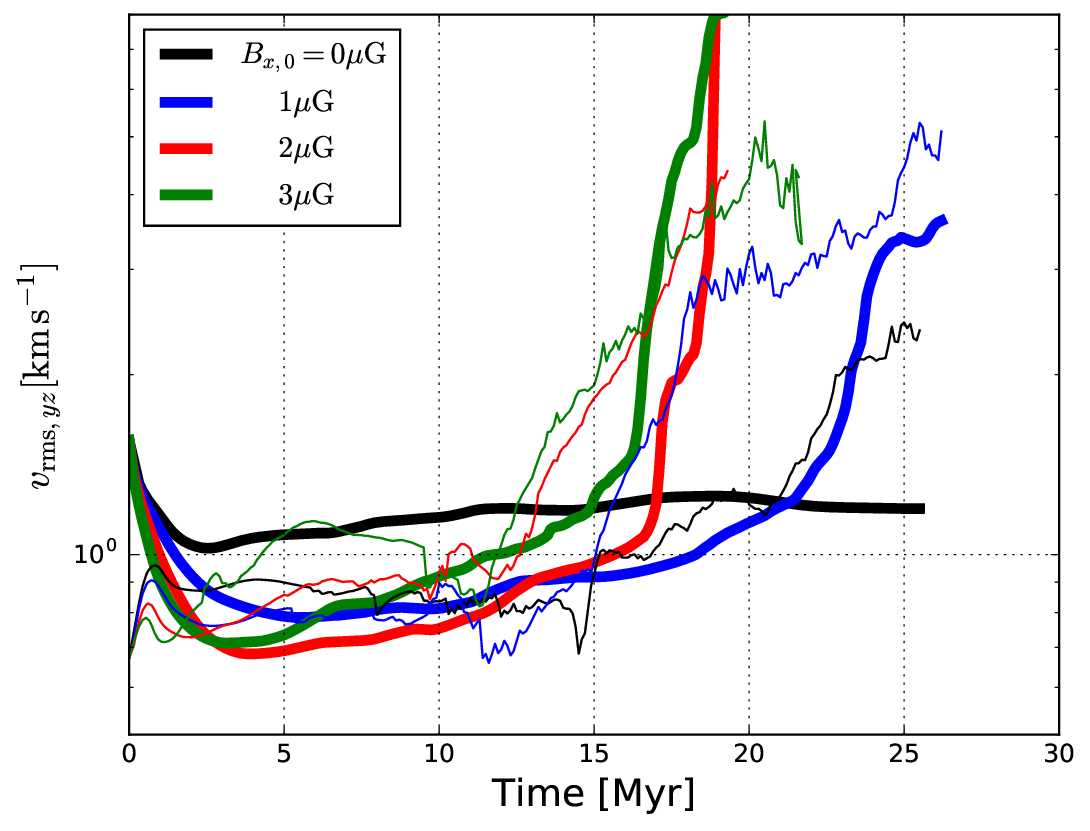}
		\includegraphics[width=0.49\hsize]{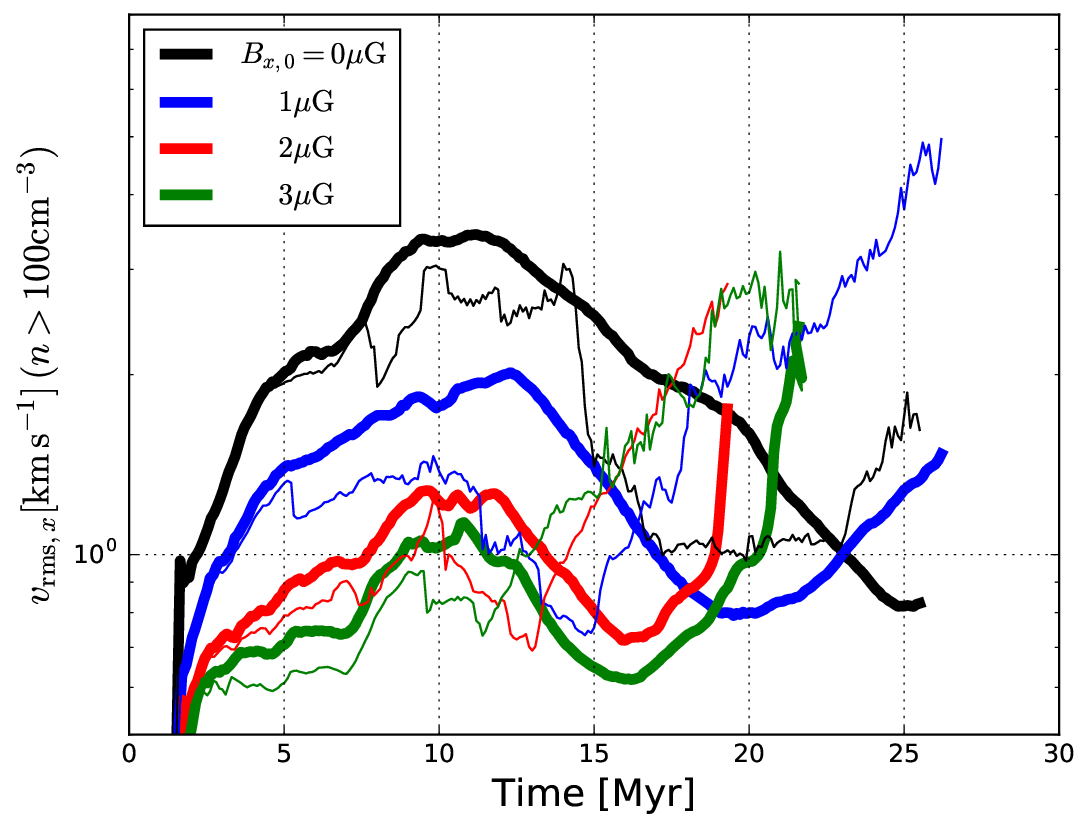}
		\caption{ {\it Left:} Evolution of the transverse velocity dispersion (in the plane perpendicular to the inflows) for all the gas in the ``central box'' (a cylinder of length and diameter equal to 80 pc, centered at the plane where the inflows collide; see Sec. \ref{sec:results}) for the 4 simulations. The thick and thin lines correspond to the velocity dispersion weighted by volume and density, respectively. {\it Right:} Evolution of the longitudinal velocity dispersion in the same box, considering only the dense gas ($n > 100 \, \ppcc$) in order to avoid picking up the inflows. The meaning of the thick and thin lines is the same as in the left panel. It is seen that both the volume- and density weighted dispersion increases for progressively smaller  magnetic field strength, being almost 4 times larger in the non-magnetic case than in the case with $B=3 \, \mu$G. The strong increase of the longitudinal (and transverse) velocity dispersion in the magnetic runs B1, B2 and B3 toward the final stages of the simulations is due to the gravitational contraction. This behavior is most evident in the density--weighted dispersion (thin lines).}
		\label{fig:transv_vdisp} 
	\end{figure*}

	As an additional test, in the {\it left panel} of Fig.\
	\ref{fig:transv_vdisp} we show the evolution of the volume--weighted velocity
	dispersion\footnote{Throughout the paper, we compute the velocity dispersion as the standard deviation of the velocity, either weighted by volume or density.}
	(thick lines)
	in the plane perpendicular to the inflows. It is seen that
	this transverse component is significantly suppressed during the early
	stages ($t \lesssim 5$ Myr) in the magnetic cases, especially in runs
	B2 and B3. As a consequence, the re-expansion motion of the compressed
	layer is inhibited, as illustrated in the {\it right panel} of Fig.\
	\ref{fig:transv_vdisp}, which shows the longitudinal (parallel to the
	inflows) velocity dispersion 
	of the dense ($n > 100 \ppcc$) gas. We
	consider the dense gas only in order to avoid including the
	longitudinal motion of the inflows. This longitudinal velocity
	dispersion is seen to be strongly suppressed during the early stages
	(again, $t \lesssim 5$ Myr) as the field strength is increased, by up
	to a factor of $\sim 4\times$ between the non-magnetic simulation B0 and
	that with $B=3\, \mu$G (B3).

	This mechanism leads to the formation of  clouds that, in the more strongly magnetized cases (runs
	B2 and B3), are denser on average, although, somewhat unexpectedly, at
	early times they contain a highly more complex filamentary structure, which is characterized by sharper,
	thinner filaments embedded in a diffuse
	background (models B2 and B3), while models B0 and B1 are
	characterized by more roundish structures and less contrast with the
	background.
	
\begin{figure*}
		\includegraphics[width=1\hsize]{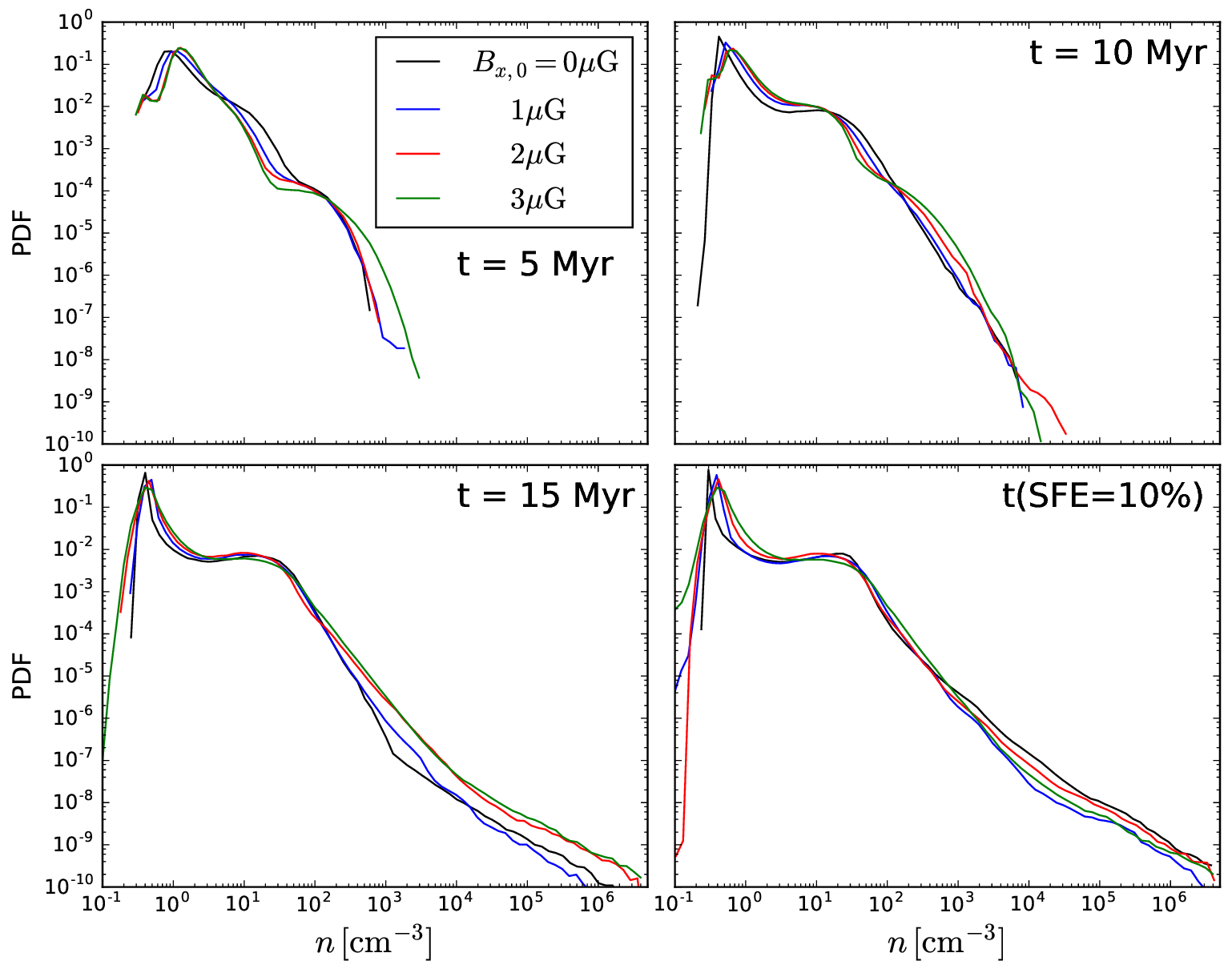}
		\caption{\label{fig:pdf} Volume-weighted density PDF for the different models at
			$t=5$, $10$, and $15 \, \Myr$ for the central box (of 80 pc per
			side). The lower right panel shows the density PDFs for all the
			models at the time at which the star formation efficiency is 10\%
			($t=25.0$, $19.8$, $16.6$, and $15.5\, \Myr$, corresponding to
			models B0, B1, B2 and B3, respectively). 
		}
	\end{figure*}

\begin{figure*}
		\includegraphics[width=1\hsize]{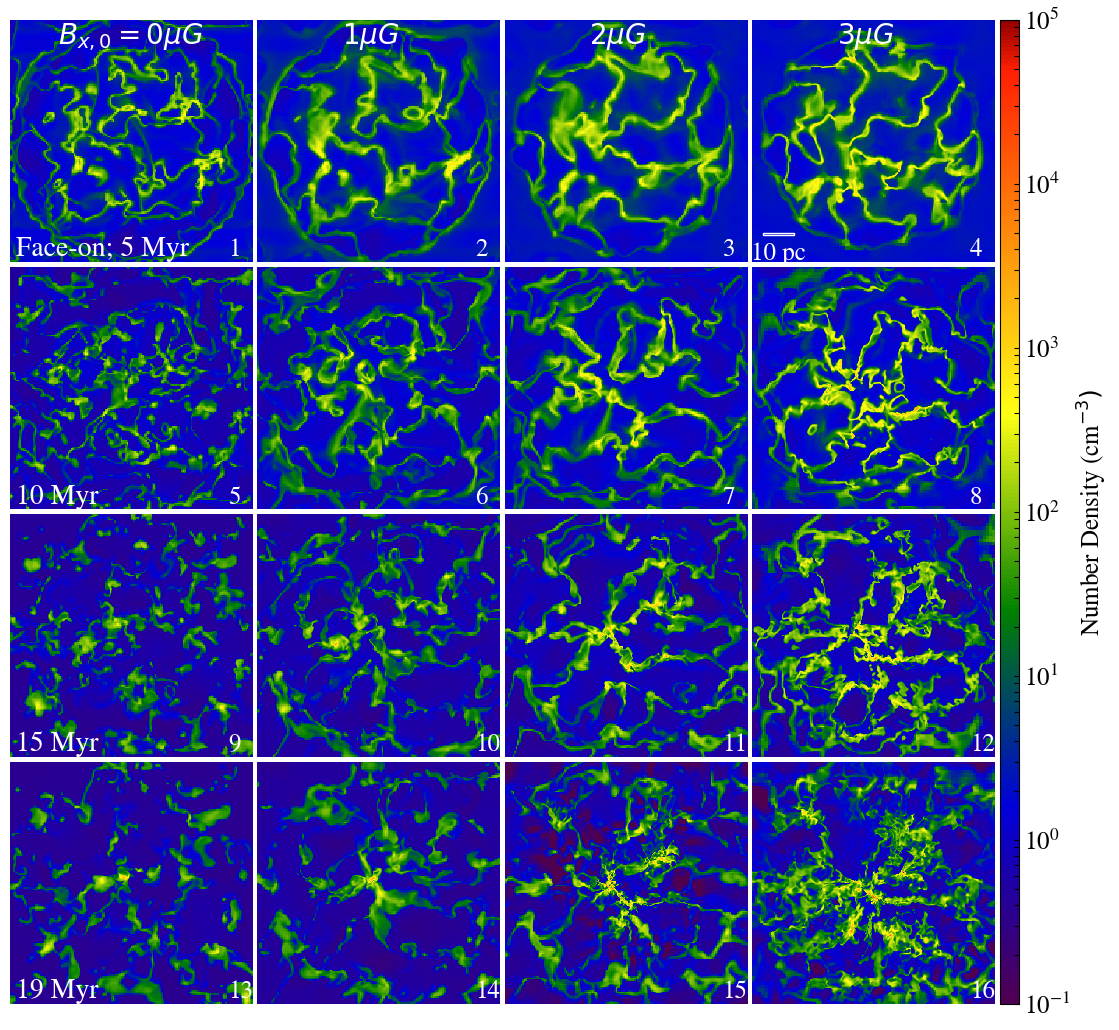}
		\caption{\label{fig:slc} Face-on slices of the number density for the ``central clouds'' at $t=5$, $10$, $15$, and
			$19 \, \Myr$. As in Fig. \ref{fig:cdens}, the different columns represent the models B0, B1, B2, and B3 and the box
			size is $80 \, \pc$ per side (see the scale in panel 4). Note that clumps and filaments are preferably green ($10\, \ppcc \lesssim n \lesssim 10^2 \ppcc$) in models B0 and B1, whereas they tend to be yellow ($10^2\, \ppcc \lesssim n \lesssim 10^3 \ppcc$) for B2 and B3 models.}
	\end{figure*}

	This suggests that the stronger NTSI, and therefore stronger 
	turbulence\footnote{Across the text, we indistinctly associate the level of turbulence with the (longitudinal) velocity
		dispersion (see, for example, the right panel of Fig. \ref{fig:transv_vdisp}).} in the less magnetized cases tends to disrupt the filaments that are produced by the TI. Note that this is a known result, \cite{SS+02} showed that the development of the TI can be attenuated (and eventually suppressed) by increasing the level of turbulent \citep[see Sec. \ref{subsec:TI}; see also][]{VS+00, AH05, VS+06}. Indeed, this is reflected in the density
	probability density function (density PDF) of the simulations, shown
	in Fig.\ \ref{fig:pdf}. In this figure, it can be seen that the more
	strongly magnetized runs (B2 and B3) tend to have a deficit of
	gas in the range $10\, \ppcc \lesssim n \lesssim 10^2 \ppcc$ and an
	excess of gas at densities $10^2 \ppcc \lesssim n \lesssim 10^3 \ppcc$,
	especially at times $t= 5$ and 10 Myr, where the dynamics is
	still not strongly dominated by gravity, but rather by the interaction
	between turbulence and TI. This is more evident in Fig. \ref{fig:slc}, which shows face-on slices at the center of the numerical box (where the inflows first collide) at different time steps for all the models. This suggests that the filaments have densities $n \sim 10^2$--$10^3\, \ppcc$, and are more abundant and sharper at
	stronger magnetic field strengths, while instead they tend to be
	destroyed by turbulence at weaker field strengths. \citet{Hennebelle13} found a similar trend, they reported that clumps formed by the TI live longer in MHD simulations compared to the ones formed in pure HD simulations. They also found that these clumps tend to be more abundant and filamentary, sharper, and denser in the MHD case.
	
	
	Interestingly, both the transverse and longitudinal density--weighted velocity dispersions shown in Fig. \ref{fig:transv_vdisp} (thin lines) exhibit a sudden increase after $t \sim 12 \, \Myr$, which we attribute to the gravitational collapse \citep[in agreement with, e.g.,][]{VS+07}. Note that the clouds with higher magnetic field strength exhibit this behavior earlier since these clouds accumulate mass faster (see Fig. \ref{fig:mgas}) and thus become gravitationally supercritical earlier.

	\subsection{Evolution of the Density PDFs}\label{subsec:PDFs}
	
	The different evolution of the clouds caused by the
	difference in the background magnetic field is also reflected on their volume-weighted
	density PDFs (see Fig. \ref{fig:pdf}). Although in general the PDFs of all four runs exhibit the
	bimodal shape characteristic of thermally bistable flows \citep[e.g.,]
	[] {VS+00, AH05, Gazol+05}, with maxima at $n \sim 0.3\, \ppcc$ and
	$30\, \ppcc$, these features do not develop clearly until $t \sim 10$ Myr.
	At earlier times ($t \sim 5$ Myr), as mentioned in Sec.\
	\ref{subsec:structure}, a prominent feature corresponding to the early
	filament formation is present at $n \sim 100\, \ppcc$, although it is
	more pronounced in cases with stronger magnetic fields, indicating that
	the stronger level of turbulence generated in the cases with weaker fields partially
	disrupts the filaments, or prevents their formation.
	
	At later times, the increasing dominance of gravity leaves its imprint
	on the PDFs. Interestingly, at $15 \, \Myr$ (see lower left panel of Fig. \ref{fig:pdf}), the density PDF for all runs exhibits two power low tails at high densities, a feature recently reported in observations toward giant molecular clouds \citep[GMCs; see e.g., ][]{Schnider+15}. 
	As time progresses, the dense gas
	fraction increases, the increase being faster in the more
	strongly magnetized cases (see this trend in the upper right and lower left panels of
	Fig.\ \ref{fig:pdf}), an effect that is directly reflected in a faster increase of the SFR (see 
	Section \ref{sec:sfe-sfr}). However, at the time at
	which the star formation efficiency (SFE; eq.\ [\ref{eq:sfe}]) is $10\%$
	in each run (at $t=25.0$, $19.8$, $16.6$, and $15.5\, \Myr$ for
	the models B0, B1, B2 and B3, respectively), the trend is reversed
	because at weaker field
	strengths the collapse takes longer to initiate, since the cloud was
	initially more dispersed. Nevertheless, at later times the collapse does
	proceed more freely for weaker field strength. This is reflected in the
	fact that run B0 takes 10 Myr more than run B3 to reach an SFE of 10\%,
	but by the time it does so, its star formation rate (SFR) is increasing
	faster (cf.\ Sec.\ \ref{sec:sfe-sfr}).
	
\begin{figure} 
		\includegraphics[width=1\hsize]{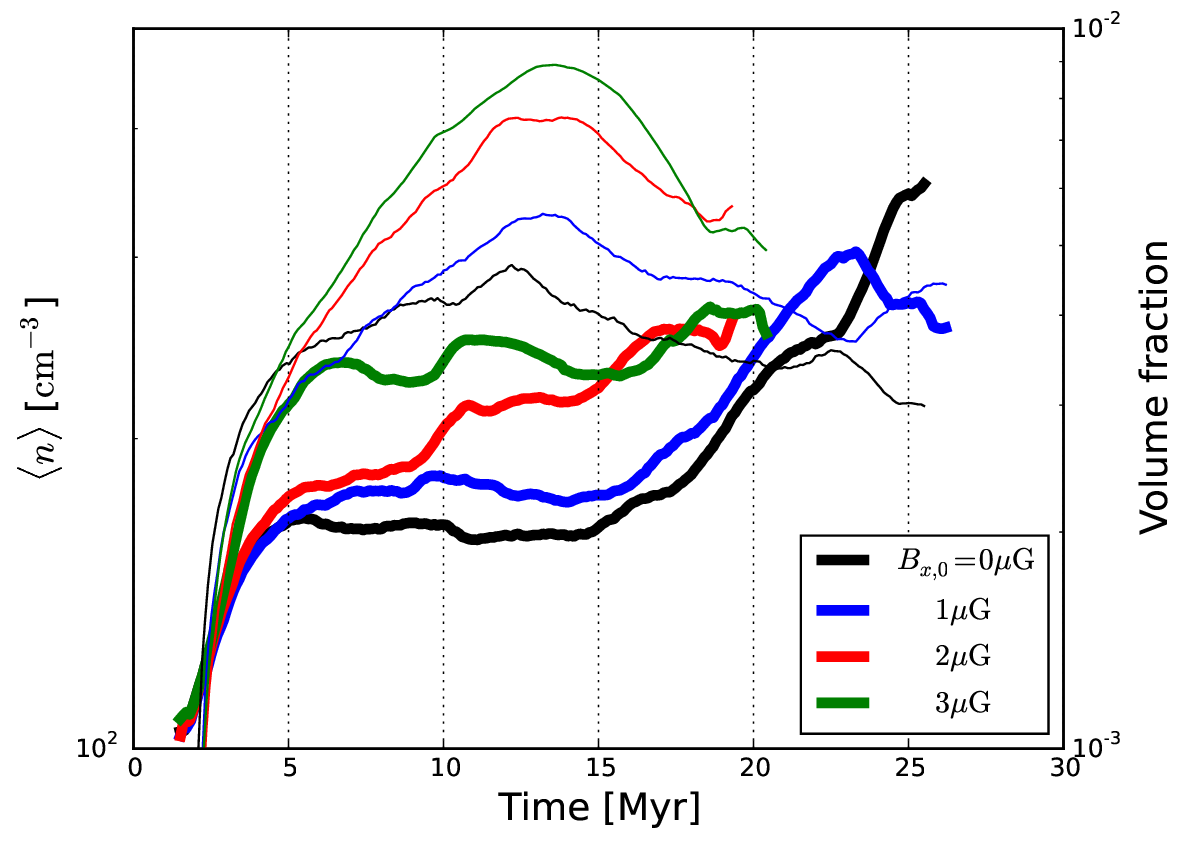}
		\caption{\label{fig:dens-vol} Evolution of the mean number density
			of dense gas ($n>100 \, \ppcc$; thick lines) of the ``central
			cloud'' in the simulations with $B_{x,0}=0$, $1$, $2$, and $3 \,
			\mG$. The thin lines represent the volume fraction of the
			dense gas in the central boxes.}
	\end{figure}

	Finally, note that, in all runs, we terminate the inflows at
	$t \sim 15$ Myr, and therefore the clouds are no longer
	fed by the diffuse gas inflows after this time. Thus, from this
	time on, the volume occupied by the dense gas ($n>100 \, \ppcc$)
	decreases while its mean density increases (Fig.\ \ref{fig:dens-vol}),
	which is a clear sign of global collapse, even for the more diffuse and
	less compact clouds (models B0 and B1).

	\subsection{Cloud mass}\label{subsec:cloud-mass}
	
\begin{figure} 
		\includegraphics[width=1\hsize]{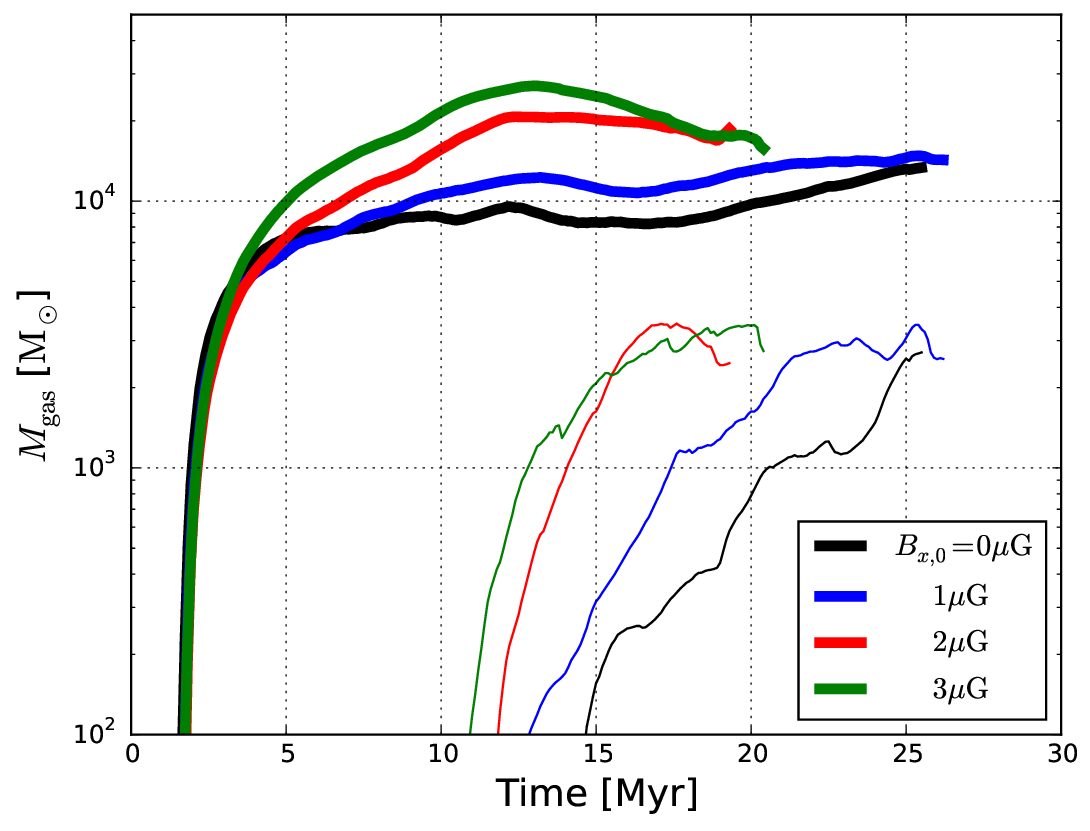}
		\caption{\label{fig:mgas} Evolution of the dense gas mass with $n>100 \, \ppcc$ (thick lines) and $n>10^{4} \, \ppcc$ (thin lines) of the ``central cloud'' for the simulations with $B_{x,0}=0$, $1$, $2$, and $3 \, \mG$. Simulations with higher (initial) magnetic field strength accumulate their mass faster and produce clouds with higher maximum masses.}
	\end{figure}

	Figure \ref{fig:mgas} shows the effect
	of increasing the magnetic field strength on the cloud mass. 
	It is seen that the clouds in the models with higher 
	magnetic field strength accumulate their mass faster and reach higher maximum masses, by up to a factor of
	$3$ (compare, for instance, models B0 and B3). This result is in 
	accordance with simulations with a similar setup by \cite{Heitsch+09}
	(see their Figure 2), but without
	self-gravity. This suggests that the stronger turbulence in the weakly magnetized
	cases counteracts the transition from the warm to the cold phase induced
	by TI, effectively reducing the amount of cold, dense gas produced.

	\subsection{The Star Formation Rate and Efficiency} \label{sec:sfe-sfr}
	
	The star formation activity in the simulations
	exhibits strong intermittency in both space and time.
	Because the clouds are contracting gravitationally,
	they are becoming denser on average.
	As shown by \citet{ZA+12}, this implies that
	quantities such as the SFE and SFR are naturally time-dependent,
	and increase in time until feedback (not included in the present
	study) begins to destroy the clouds.
	
	As in previous works \citep[e.g.,] [] {VS+07, VS+10,
		ZA+12, Colin+13, ZV14}, we define the instantaneous SFE as
	\beq \label{eq:sfe}
	{\rm SFE}(t) = \frac{\Ms(t)}{\Mc(t) + \Ms(t)},
	\eeq
	where $\Mc$ is the cloud mass (specifically, the
	gas mass at $n>100 \, \ppcc$), and $\Ms$
	is the total mass in stars, which we take
	as the total mass in sinks, $\Msinks$.
	
\begin{figure}
		\includegraphics[width=1\hsize]{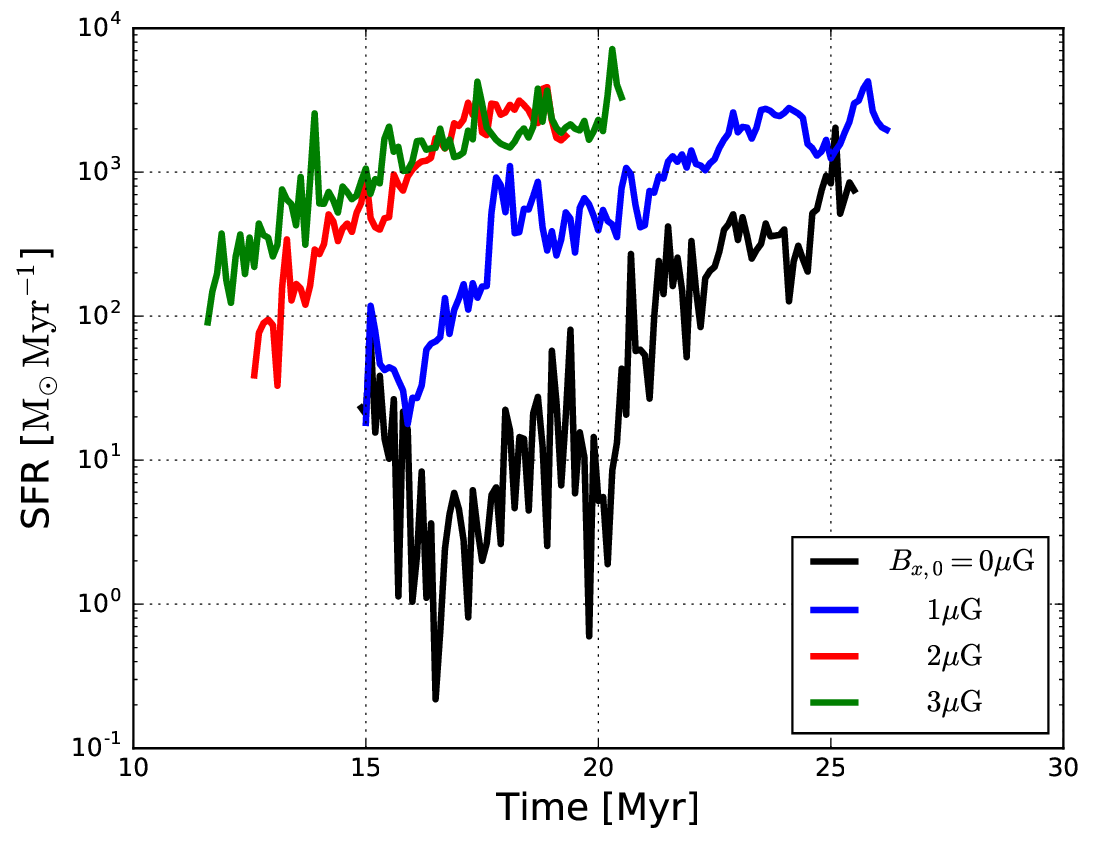}
		\caption{\label{fig:sfr} Evolution of the SFR. The onset of SF is seen to be delayed as the
			background magnetic field strength is decreased, although it is
			nearly simultaneous for runs B1 and B0.}
	\end{figure}

\begin{figure}
		\includegraphics[width=1\hsize]{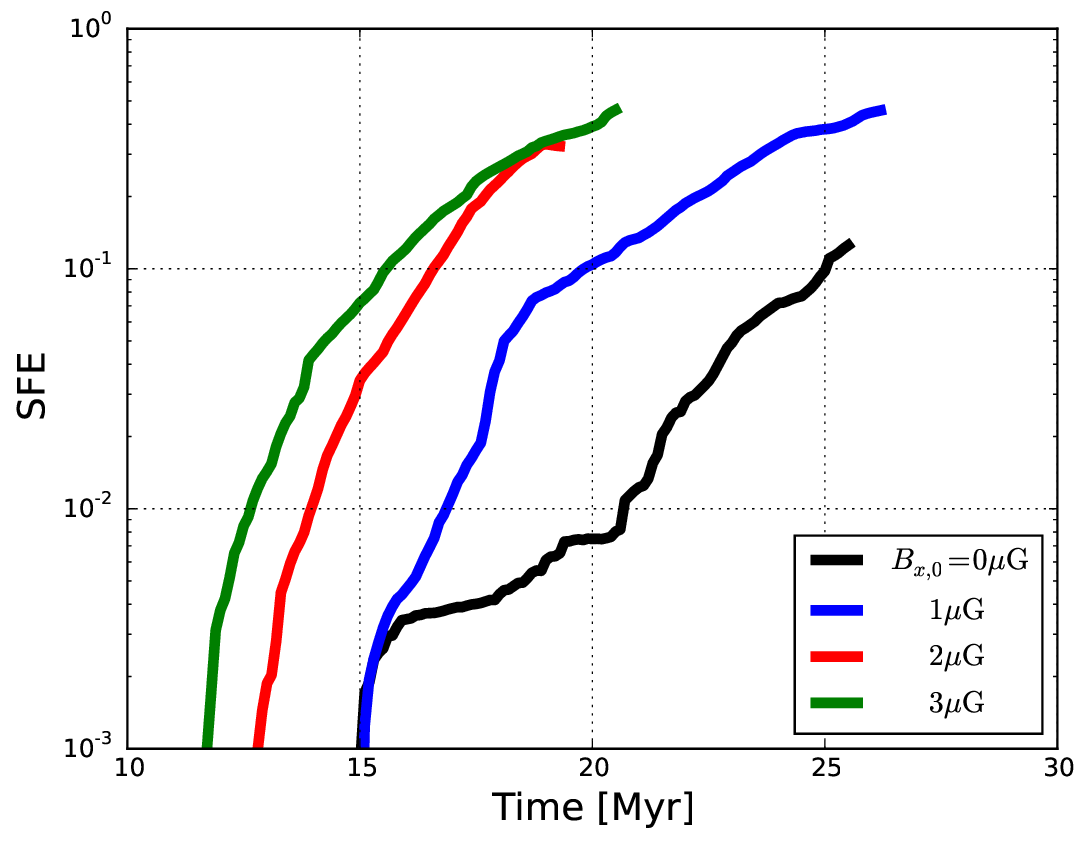}
		\caption{\label{fig:sfe} Evolution of the star
			formation efficiency (SFE; see eq.\ \ref{eq:sfe}) for
			the four simulations. Note that, because stellar feedback is not
			included (cf.\ Sec. \ref{sec:heating-cooling}), the SFE is seen to
			reach values larger than observed values for whole GMCs \citep[see e.g.,][]{VS+10,Colin+13,Dale+13,Walch+15a}).}
	\end{figure}

	We estimate the SFR as the time derivative of the
	total sink mass, $d \Msinks (t)/ d t$ (which is calculated by dividing
	the difference in the total sink mass in a time step by the duration of
	the step). This accounts for
	both the mass collapsed onto new sinks and the mass accreted onto the
	existing ones. Figures \ref{fig:sfr} and \ref{fig:sfe} show the
	time evolution of the SFR and SFE for the various simulations
	with different initial strengths of magnetic field.  In general,
	it is seen that both the SFE and the SFR increase monotonically in
	time until the end of the simulations. This is due to the inherent
	increase of the SFR caused by the collapse of the clouds, and to the neglect
	of stellar feedback (cf.\ Sec.\ \ref{sec:heating-cooling}), which has
	been shown previously to partially or completely suppress the SF activity in
	the clouds, depending on their mass (and possibly geometry)
	\citep[e.g.,] [] {VS+10, Dale+12, Dale+13, Colin+13}. Therefore, in the
	present simulations, there is no agent that prevents the continued
	increase of the SFR. However, this allows us to investigate the effects
	of varying the magnetic field alone.
	
	In this context, several points are worth noting in Figs.\
	\ref{fig:sfr} and \ref{fig:sfe}.  First, it is seen that the onset of
	SF is delayed by larger amounts at weaker values of the magnetic field
	strength, although it is nearly simultaneous for runs B1 and B0.
	Morever, while run B2 and B3 start forming sinks {\it before} the
	inflows end ($t \sim 15$ Myr), runs B0 and B1 only begin forming sinks
	when the inflows have ended. Second, run B0 exhibits an initial period
	of nearly 2 Myr during which the SFR {\it decreases}, but after that
	time, its SFR begins increasing. 
	A similar initial decrease of the SFR is
	observed in run B1, although its duration is shorter,
	$\sim 1$ Myr.  Third, it is seen that, once the SFRs of all
	simulations are increasing, the SFR increase rate of run B0 is larger
	than that of the magnetic runs (see Section \ref{subsec:PDFs}).  Finally, note that the SFE (as well
	as the SFR) is much higher than that typically observed in GMCs
	\citep[$\sim 2$\%, e.g.,][]{Myers+86}
	due to the absence of stellar feedback (see Secs.\
	\ref{sec:heating-cooling} and \ref{sec:missingphysics}). However, this
	is not a problem because here we are not attempting to reproduce
	actual observed SFEs, but rather to isolate the effects of varying
	the magnetic field on the SFR and the SFE.

	\begin{figure}
		\includegraphics[width=1\hsize]{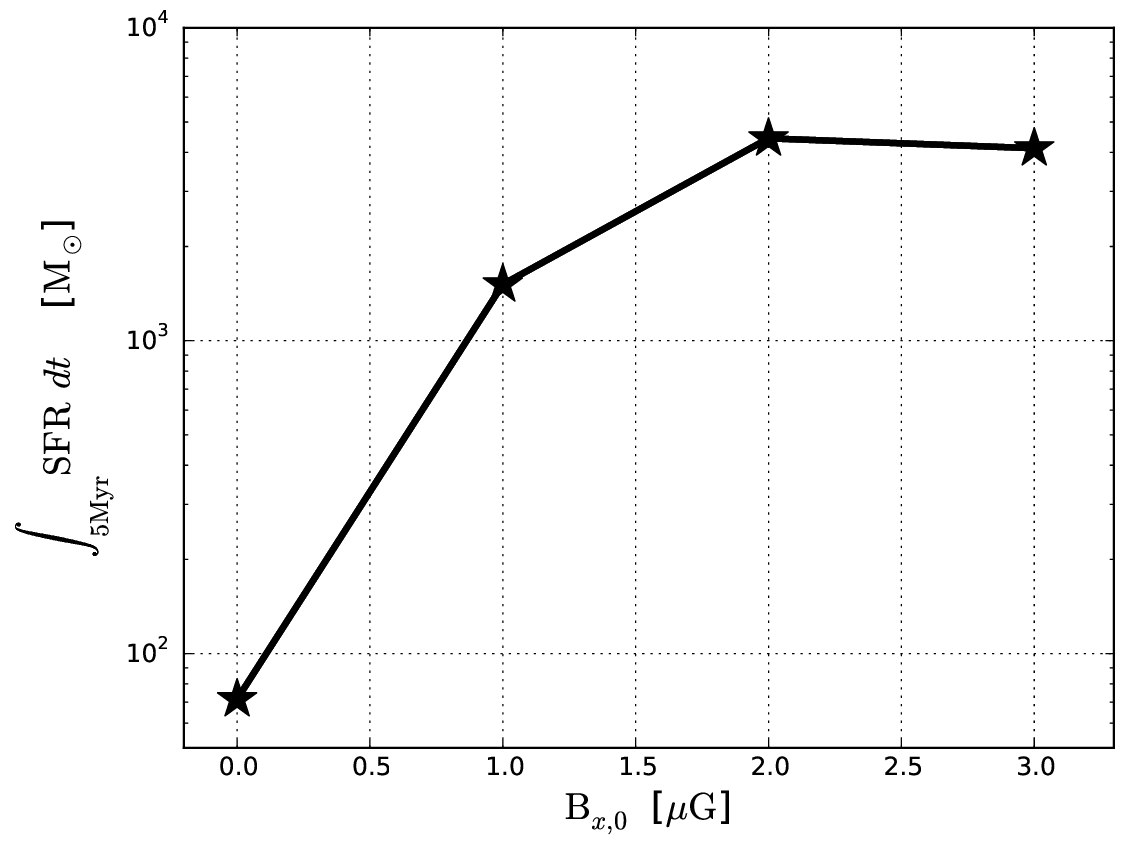}
		\caption{\label{fig:MsB} Total mass in sinks $5 \, \Myr$ after the onset
			of star formation (i.e., the SFR integrated during this time interval)
			for all runs. Note that, on average, the star formation activity
			increases with increasing magnetic field strength.}
	\end{figure}

	\section{Discussion}\label{sec:discussion}
	
	\subsection{Comparison with previous work}
	
	The effect of the magnetic field strength and
	orientation in the assembly and star-forming activity of MCs has been
	addressed by several studies in the past \citep[see
	e.g.,][]{Passot+95, HP00, BP-MacLow02, Heitsch+09, Banerjee+09, VS+11,
		Hill+12, Walch+15b, Bastian-Robi15}. In particular, \cite{Passot+95}
	showed that the dependence of the SFR on the magnetic field
	strength is highly non-monotonic. In
	particular, at intermediate magnetic field strengths, an
	increase of the uniform background field tends to suppress shearing
	motions in the medium that disrupt clouds,
	thereby promoting the
	growth of density fluctuations \citep[see
	also][]{BP-MacLow02}. In our case, the shearing motions are caused
	by the NTSI, and their inhibition by the magnetic field leads to the
	formation of more massive clouds (cf.\ Sec.\ \ref{subsec:cloud-mass}
	and Fig.\ \ref{fig:mgas}) and 
	therefore a larger total stellar mass (Fig.\ \ref{fig:MsB}).
	The latter
	shows the total mass in sinks 5 Myr after the onset of star formation
	in all runs, and can be compared to Figure 8 in \citet{Passot+95}.
	Note that although the setup of those authors
	was very different from
	ours (as well as the numerical techniques), and the shearing in
	the study of those authors was due to differential galactic
	rotation, not the NTSI, the effect of magnetic fields on reducing shearing motions that inhibit contraction is similar.
	
	In addition, our results are consistent
	with those of \citet{Heitsch+09},  who
	also studied the problem of cloud formation by
	converging flows in the case where the magnetic fields are aligned
	with the flows, although they
	considered only the role of the magnetic
	field in the process of cloud formation, excluding self-gravity
	\citep[see also][]{Heitsch+07}. Those authors found that
	increasing
	the magnetic field strength tends to
	weaken (or even suppress) the NTSI, which results in clouds with
	reduced levels of turbulence, larger
	masses and smaller sizes, in agreement
	with our results (see Fig.\ \ref{fig:mgas}).

	Our findings can also be compared to those of
	\citet[][hereafter CN+14]{CN+14}, who used a setup 
	similar to ours, but not including the magnetic field. Interestingly,
	rather than investigating the effect of variations of the
	intensity of the background magnetic field, those authors were concerned with
	the effect of pre-existing clumpiness in the colliding flows. They found
	that clouds formed by
	clumpy inflows are in general more turbulent and
	have lower total mass (gas + sinks) as well as fewer but more
	massive sink particles. Therefore, the
	cloud formed by clumpy inflows in that study is similar to the clouds
	formed at lower field strengths in our case, although for a different
	reason: in the case of CN+14, the cloud is scattered because the
	clumpiness of the incoming flows reduces the amount of mass that collides
	directly against material from the opposing inflow, thus reducing the
	dissipation of the kinetic energy of the inflows. Instead, in our case,
	the presence of progressively stronger magnetic fields causes a
	suppression of the NTSI, reducing the generation of turbulence in the
	cloud, as well as the ``bounce'' (i.e., the rebound of the dense layer because of an increase in its internal turbulent ram pressure; see Fig. \ref{fig:pram_pmag}) that results from the development of
	the instability (see Sec.\ \ref{subsec:structure}).
	
	It is worth noting that we use a very similar setup as that in the works presented by \citet[][hereafter B+09, VS+11, and KB15, respectively]{Banerjee+09,VS+11,Bastian-Robi15}, although with different parameters in the initial conditions as shown in Table \ref{tab:pwork}. Our model B2 ($\mu / \mucrit = 2.38$) can be compared with the simulation presented by \cite{Banerjee+09} ($\mu / \mucrit \approx 2.9$). 
	Comparing the SFE at the end of the simulations (see their Fig. 13 and Fig. \ref{eq:sfe} in this work) we find quite similar values in both models (around $\sim$40-50\%).
	
	On the other hand, VS+11 present simulations with $\mu / \mucrit \approx 1.36, 0.91$, and 0.68, so we can compare their magnetically supercritical simulation (B2-AD model) with our model B3 ($\mu / \mucrit \approx 1.59$). Their SFE saturates again at $\sim 40$\%, in agreement with our B3 model (compare the lower-right panel of their Fig. 5 to our Fig. \ref{fig:sfe}). 
	Note that the evolving time scales are different in the mentioned works since the inflow parameters are different. It also should be noticed, however, that most of the inflow parameters used in this work are different to those used in \citet{Banerjee+09} and VS+11 (see Table \ref{tab:pwork}) suggesting that the dominant parameter controlling the turbulence level and the SF activity itself is the ``magnetic criticality'' (through the $\mu / \mucrit$ ratio), although more research is needed in this regard (particularly an uniform parameter study).
	
	Lastly, KB15 
	only studied magnetically subcritical clouds and so our models cannot be compared directly, but can be considered complementary to our work instead. Interestingly, these authors find that in sub-critical clouds (specifically in their models with $\mu / \mucrit \approx$ 0.59 and 0.47 listed in Table \ref{tab:pwork}) the NTSI may be suppressed since the $\beta_{\rm ram}$ parameter is less than 1 (following the trend shown in Fig. \ref{fig:pram_pmag}; see also Fig. 12 in KB15).

	\subsection{Test of the refinement criterion} \label{subsec:ref} 
	
	Our {\it constant-mass} refinement criterion does not fulfill the
	Truelove criterion \citep{Truelove+97} for
	prevention of spurious
	fragmentation. This criterion requires the resolution to scale linearly with the
	Jeans length $\LJ$ (i.e., as $\rho^{-1/2}$, assuming constant
	temperature; see eq.\ [\ref{eq:lJ}]). Instead, with our criterion, the resolution scales as
	$\rho^{-1/3}$, or equivalently, as $\LJ^{2/3}$, implying that the Jeans
	length is increasingly more poorly resolved as the density increases, and therefore, some artificial 
	fragmentation may be expected. 
	
	\begin{figure}
		\includegraphics[width=1\hsize]{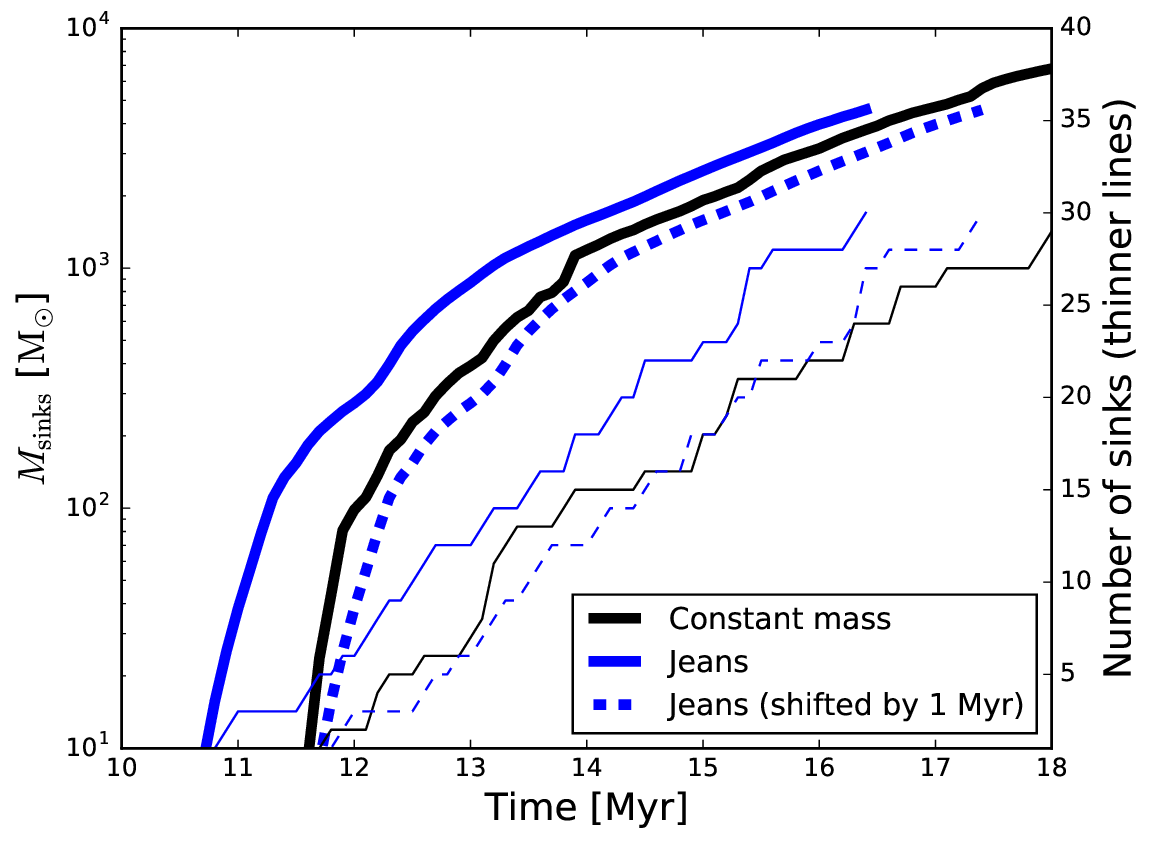}
		\caption{\label{fig:ref} Time evolution of the number (thin lines) and the total mass in sinks (thick lines) for runs 
			B3 (which uses the {\it constant mass} refinement criterion) 
			and B3J (Jeans criterion). Note that both the number and the total mass in sinks are very similar when we consider the curves for the run B3J shifted by $1 \, Myr$ (discontinuous lines).}
	\end{figure}
	
	In order to quantify this effect, we performed 
	an additional simulation, labeled B3J, which is
	identical to run B3 except for the refinement criterion used. For run
	B3J we use the Jeans criterion and we resolve the Jeans length (eq. \ref{eq:lJ}) 
	by at least 10
	grid cells in order to fulfill the Truelove criterion
	; i.e., the condition $\lambda_{\rm J} \geq 10 \times \Delta x$ is satisfied in every cell.
	At the higher level of refinement, this condition imposes a density threshold for sink formation of $\rhoth = 7.1 \times 10^{-20} \, \gpcc$($= 1.8 \times 10^4 \,
	\ppcc$), which is two orders of magnitude lower than the threshold used in the {\it constant-mass criterion}. Thus, using the Jeans criterion instead of the {\it constant-mass} one, we expect an earlier onset of star formation.
	Figure \ref{fig:ref} shows the evolution of the
	number of sink particles (thin continuous lines; right $y$-label) and the total
	mass in them (thick continuous lines). It can be seen that both
	quantities are very close for the two simulations, and
	that the effect of choosing the
	Jeans criterion over the {\it constant mass} criterion is just to
	delay the onset of star formation by $10$\%, as can be appreciated from the dashed lines in the same figure, which were shifted by 1 Myr just to illustrate this difference. Therefore, the effect of the
	refinement criterion on our results is just an offset in the onset of star formation. For instance, a shift of 1 Myr puts run B3 in the same temporal starting point as run B2 in Fig. \ref{fig:sfr}. However, we expect the rest of the cloud properties to remain qualitatively the same.

	\subsection{Limitations}
	
	\subsubsection{Neglected processes}
	\label{sec:missingphysics} 
	
	One of the most notable
	omissions in our simulations is that of stellar
	feedback. In particular, the ionizing 
	radiation from massive
	stars is generally considered to be the dominant mechanism of stellar
	energy injection into GMCs \citep{Matzner02}, 
	although its effect may be mostly the dispersal of
	the clouds rather than maintaining them in a
	near-equilibrium state and reducing both the SFR and the SFE
	\citep{VS+10, ZA+12, Dale+12, Dale+13, Colin+13}.  Moreover, after a
	few Myr, SN explosions are expected to begin occurring, contributing to
	further cloud dispersal, although the effect may not be stronger than
	that of ionizing radiation \citep[e.g.,] [] {Walch+15a, IH15,Bastian+16}.
	
	Nevertheless, as mentioned above, in this paper we have not been
	concerned so much about generating a realistic model of MCs at all times
	in their evolution, but rather in performing a series of numerical
	experiments assessing the effect of the uniform magnetic field strength
	in setting up the initial conditions of the clouds. For this purpose,
	our numerical simulations are appropriate.
	
	%

	\section{Summary and Conclusions} \label{sec:conclusions}
	
	In this paper, we have presented four MHD simulations aimed
	at studying the evolution
	of MCs formed by diffuse converging flows in the presence of magnetic
	fields, in order to investigate the effect of the
	latter on the development of 
	the initial conditions of the nascent clouds and their early star
	formation activity. We have considered four otherwise identical numerical
	models except for the initial magnetic 
	field strength, for which we have considered the values $0$,
	$1$, $2$, and $3 \, \mG$ (models B0, B1, B2, and B3, respectively),
	corresponding to magnetically supercritical configurations in all cases.
	The field is initially aligned with the flows.
	The latter is not a restrictive condition, however, since it has been
	shown that, when the flows are oblique to the magnetic field and have
	sufficiently high Mach numbers (depending on the field strength), the
	flows are reoriented to proceed along the magnetic field \citep{HP00}.
	Therefore, whenever the flows manage to collide, they will do so aligned
	with the field.
	
	In the initial phase of cloud formation, our simulations show
	that, in agreement with the non-self-gravitating results from
	\citet{Heitsch+09}, the presence of a magnetic field aligned with the
	inflows tends to suppress the NTSI, causing the clouds to be more
	compact, denser, and less turbulent. In the presence of self-gravity,
	this leads to an earlier global collapse of the clouds, and to an
	enhanced early star-formation activity. Instead, more weakly magnetized,
	or fully non-magnetic cases produce more dispersed and more turbulent
	clouds, that initially expand in a rebound from the initial collision,
	and only later begin to collapse. However, when they do so, their SFRs
	increase more rapidly, because the collapse is then unimpeded by the
	magnetic field. 
	
	Moreover, the non-magnetic simulation, run B0, and that with the weakest
	magnetic field (B1), do not start their collapse until the inflows are
	terminated.
	Note, however, that this does not necessarily imply that the turbulence generated by the inflows is sufficient to support the clouds, since it is possible that, if the inflows were maintained for longer times, the clouds could gain enough mass  to begin their collapse (further study is necessary in order to address this issue).
	Nevertheless, because the accepted value of the mean magnetic field
	strength in the ISM is comparable to that considered in our more
	strongly magnetized cases \citep[runs B2 and B3;] [] {Beck+96}, our
	results reinforce the notion that the turbulence produced by the stream
	collision in the nascent clouds is not enough to support them against
	their self-gravity \citep{VS+07}, and therefore engage in global
	gravitational contraction soon after they are assembled. 
	
	This conclusion is further
	reinforced by the fact that the clouds in runs B0 and B1, which {\it
		are} supported by turbulence in their initial stages, do not begin
	forming stars until {\it after} the inflows, and therefore the
	turbulence driving, are terminated, and the clouds engage in global
	gravitational contraction. This suggests that {\it star-forming} GMCs
	are in general in a state of global collapse.

	\section*{Acknowledgements}

	We thank the referee for helpful and constructive comments that improved the clarity of the paper. We thankfully acknowledge Fabian Heitsch for sti\-mu\-la\-ting discussions. MZA acknowledges hospitality offered by Robi Banerjee and Ralf Klessen at {Hamburg Observatory} (University of Hamburg) and {Institute of Theoretical Astrophysics} (University of Heidelberg) during the first stages of this work, and financial support from PAPIIT grants IN111313 and IN110214, CONACyT grant CB152913, and CONACyT postdoctoral fellowship at University of Michigan. EV-S acknowledges financial support from CONACYT grant 255295. RB and BK acknowledge funding from the German Science Foundation (DFG) for this project via the ISM-SPP 1573 grant BA 3706/3-1 and BA 3706/3-2. RB also acknowledge funding from the DFG via the grant BA 3706/4-1. LH was supported in part by NASA grant NNX16AB46G and by the University of Michigan. We also acknowledge Christopher Davies, Gilberto Zavala P{\'e}rez, Alfonso H. Ginori Gonz{\'a}lez, and Miguel Espejel Cruz for their valuable computational support. The numerical simulations presented here were performed on the {\it Calzonzin} cluster at {Instituto de Radioastronom{\'\i}a y Astrof{\'\i}sica} (Universidad Nacional Aut{\'o}noma M{\'e}xico), acquired through the CONACYT grant 102488 to EVS. The visualisation was carried out with the {\tt yt} software \citep{yt}. The FLASH code used in this work was in part developed by the DOE NNSA-ASC OASCR Flash Center at the University of Chicago.This research has made use of NASA's Astrophysics Data System Abstract Service.

\bibliographystyle{mnras}
\bibliography{refs}

%
%
%
\bsp	
\label{lastpage}
\end{document}